\documentclass[fleqn,usenatbib]{mnras}

\usepackage[T1]{fontenc}

\DeclareRobustCommand{\VAN}[3]{#2}
\let\VANthebibliography\thebibliography
\def\thebibliography{\DeclareRobustCommand{\VAN}[3]{##3}\VANthebibliography}

\usepackage{xcolor}
\usepackage{listings}
\usepackage{booktabs}

\usepackage{graphicx}	
\usepackage{amsmath}	
\usepackage{amssymb}	

\usepackage{newtxtext,newtxmath}

\usepackage{ulem}
\newcommand{\strike}[1]{}
\newcommand{\revise}[1]{#1}

\title[Asteroseismic Fingerprints of Stellar Mergers]{Asteroseismic Fingerprints of Stellar Mergers}

\author[N. Z. Rui and J. Fuller]{
Nicholas Z. Rui,$^{1}$\thanks{E-mail: nrui@caltech.edu}
Jim Fuller$^{1}$
\\
$^{1}$TAPIR, California Institute of Technology, Pasadena, CA 91125, USA}

\date{Accepted XXX. Received YYY; in original form ZZZ}

\pubyear{2021}

\begin{document}
\label{firstpage}
\pagerange{\pageref{firstpage}--\pageref{lastpage}}
\maketitle

\begin{abstract}
Stellar mergers are important processes in stellar evolution, dynamics, and transient science.
However, it is difficult to identify merger remnant stars because they cannot easily be distinguished from single stars based on their surface properties.
We demonstrate that merger remnants can potentially be identified through asteroseismology of red giant stars using measurements of the gravity mode period spacing together with the asteroseismic mass.
For mergers that occur after the formation of a degenerate core, remnant stars have over-massive envelopes relative to their cores, which is manifested asteroseismically by a g~mode period spacing smaller than expected for the star's mass.
Remnants of mergers which occur when the primary is still on the main sequence or whose total mass is less than $\approx\! 2 \, M_\odot$ are much harder to distinguish from single stars.
Using the red giant asteroseismic catalogs of \citet{vrard2016period} and \citet{yu2018asteroseismology}, we identify \revise{$24$}\strike{$15$} promising candidates for merger remnant stars.
In some cases, merger remnants could also be detectable using only their temperature, luminosity, and asteroseismic mass, a technique that could be applied to a larger population of red giants without a reliable period spacing measurement.
\end{abstract}




\section{Introduction} \label{sec:intro}

Stellar mergers are physically complex processes with broad implications across astrophysics.
\citet{kochanek2014stellar} find that galactic mergers
occur at a high rate of $\sim \! 0.2$ yr$^{-1}$, and
\citet{de2014incidence} further show that merger products comprise $\approx \! 30\%$ of high-mass main sequence stars.
Mergers are a common endpoint of binary stellar evolution \citep{paczynski1976common}, and they are believed to be the origin of astrophysical transients such as \strike{such as }luminous red novae \citep{tylenda2006eruptions,soker2006violent,ivanova2013identification,pejcha2016binary,metzger2017merger}.
Collisions between stars are also expected to occur at high rates in dense stellar environments such as globular clusters, where they are believed to be an important formation channel for blue stragglers \citep{bailyn1995blue}, some of which have exotic properties suggestive of this origin \citep[e.g.,][]{schneider2019stellar}.
In these environments, they have been recently proposed as one possible explanation for multiple stellar populations \citep{mastrobuono2019mergers,wang2020possible}.

\begin{figure*}
    \centering
    \includegraphics[width=\textwidth]{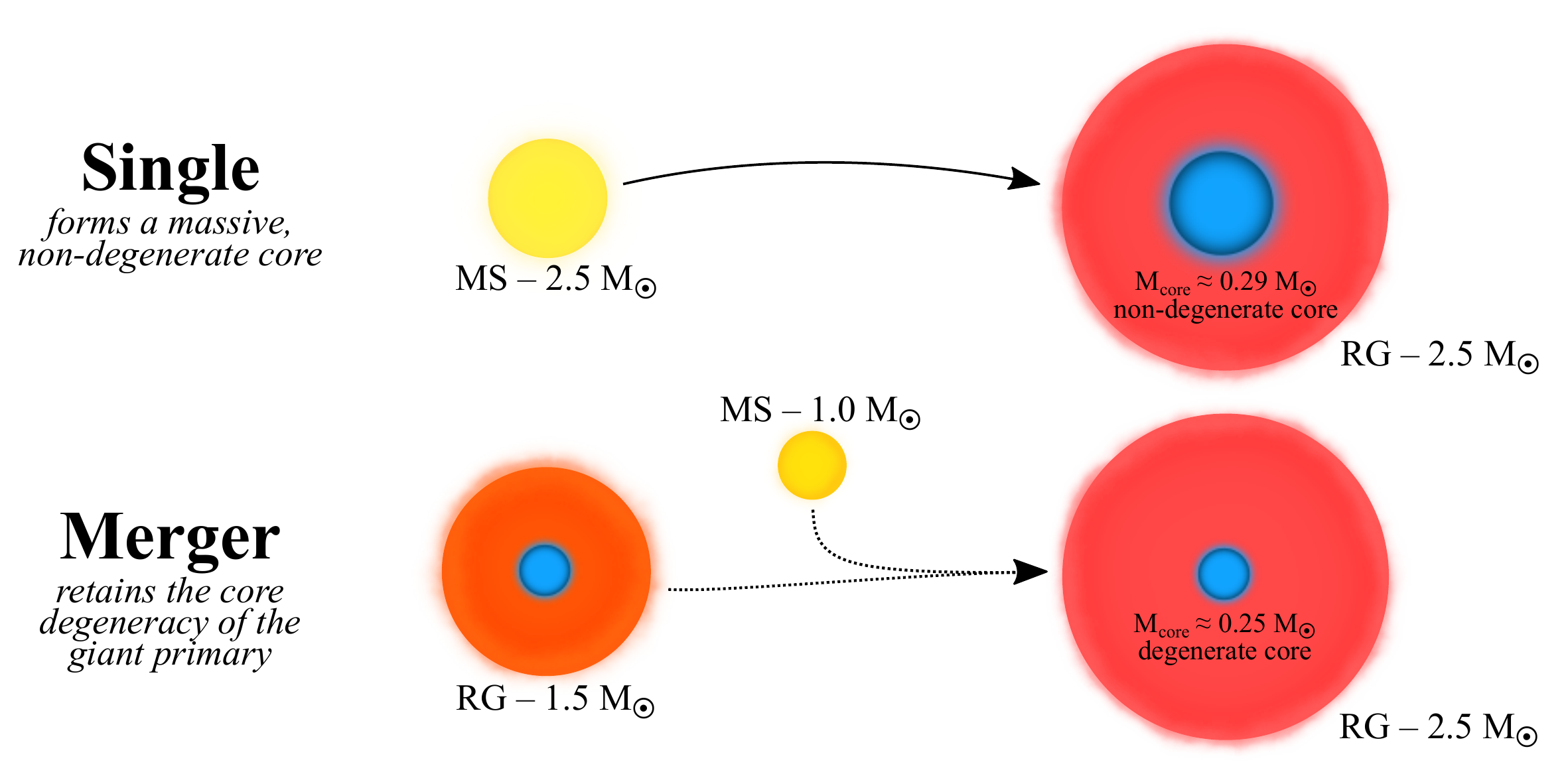}
    \caption{A cartoon comparing a $2.5$ $M_\odot$ RG formed as an isolated, single star with one formed in a merger where the primary has already entered the RG phase and the secondary remains on the MS.
    If the RG primary is in the mass range where it forms a degenerate core (e.g., $M=1.5$ $M_\odot$), the merger product will generally retain it, even if a single star of the same mass would have been expected to produce a more massive, non-degenerate core on the RGB.}
    \label{fig:merger_cartoon}
\end{figure*}

While millions of merger remnants are expected to exist in the Galaxy, identifying the surviving stars in the field is challenging.
Detailed asteroseismic characterization of red giant (RG) stars offers a new hope, because the oscillations of RGs are particularly rich in information for two reasons.
First, the close values of the Brunt--V\"ais\"al\"a and Lamb frequencies create a narrow evanescent region within the star, coupling the observable p~modes at their surfaces to the g~modes within their radiative cores.
Second, the frequencies occupied by these ``mixed modes'' are serendipitously excited by stochastic driving from convective motions in their envelopes.
The intimate coupling between interior and surface oscillations allow for detailed asteroseismic constraints on their core structures, allowing for the determination of evolutionary states \citep{bedding2011gravitymodes,bildsten2011acoustic,mosser2014mixed,cunha2015structural,elsworth2017new}, internal rotation rates \citep{beck2012fast,mosser2012rotation,klion2017diagnostic,gehan2018core,ahlborn2020asteroseismic,deheuvels2020seismic}, and core magnetic fields \citep{fuller2015asteroseismology,stello2016suppression,cantiello2016asteroseismic,mosser2017dipole,loi2020magnetic}.

Therefore, in addition to encoding stellar masses and radii in the large frequency spacing $\Delta\nu$ and the frequency of maximum power $\nu_{\mathrm{max}}$ \citep{kallinger2010oscillating}, asteroseismology is also a probe of the core structures of RGs through the mixed mode period \strike{splitting}\revise{spacing} $\Delta P_g$ in the dipole ($\ell=1$) mode peaks of approximately
\begin{equation} \label{eqn:gravity}
    \Delta P_g = \sqrt{2}\pi^2\left( \int_{\mathcal{R}} \frac{N}{r} \mathrm{d}r \right)^{-1}\mathrm{.}
\end{equation}
Here, $N$ is the Brunt--V\"ais\"al\"a frequency and $\mathcal{R}$ denotes the portion of the star's central radiative region where $\nu_{\mathrm{max}}<N$ \citep{chaplin2013asteroseismology}.
Intuitively, this dipole splitting results from the coupling of a given p~mode to multiple, distinct g~modes, hence the sensitive dependence of $\Delta P_g$ on the Brunt--V\"ais\"al\"a frequency in the central regions of the star.
For a typical RG, $\Delta P_g$ is typically on the order of minutes, large enough to be measured by prominent surveys such as \textit{CoRoT} \citep{mosser2011mixed} and \textit{Kepler} \citep{stello2013asteroseismic,vrard2016period}.
The value of $\Delta P_g$ is primarily determined by the mass and evolutionary state of the star's helium core, and comparison of measured values to models provides an excellent test of stellar evolution theories. 

\begin{figure*}
    \centering
    \includegraphics[width=\textwidth]{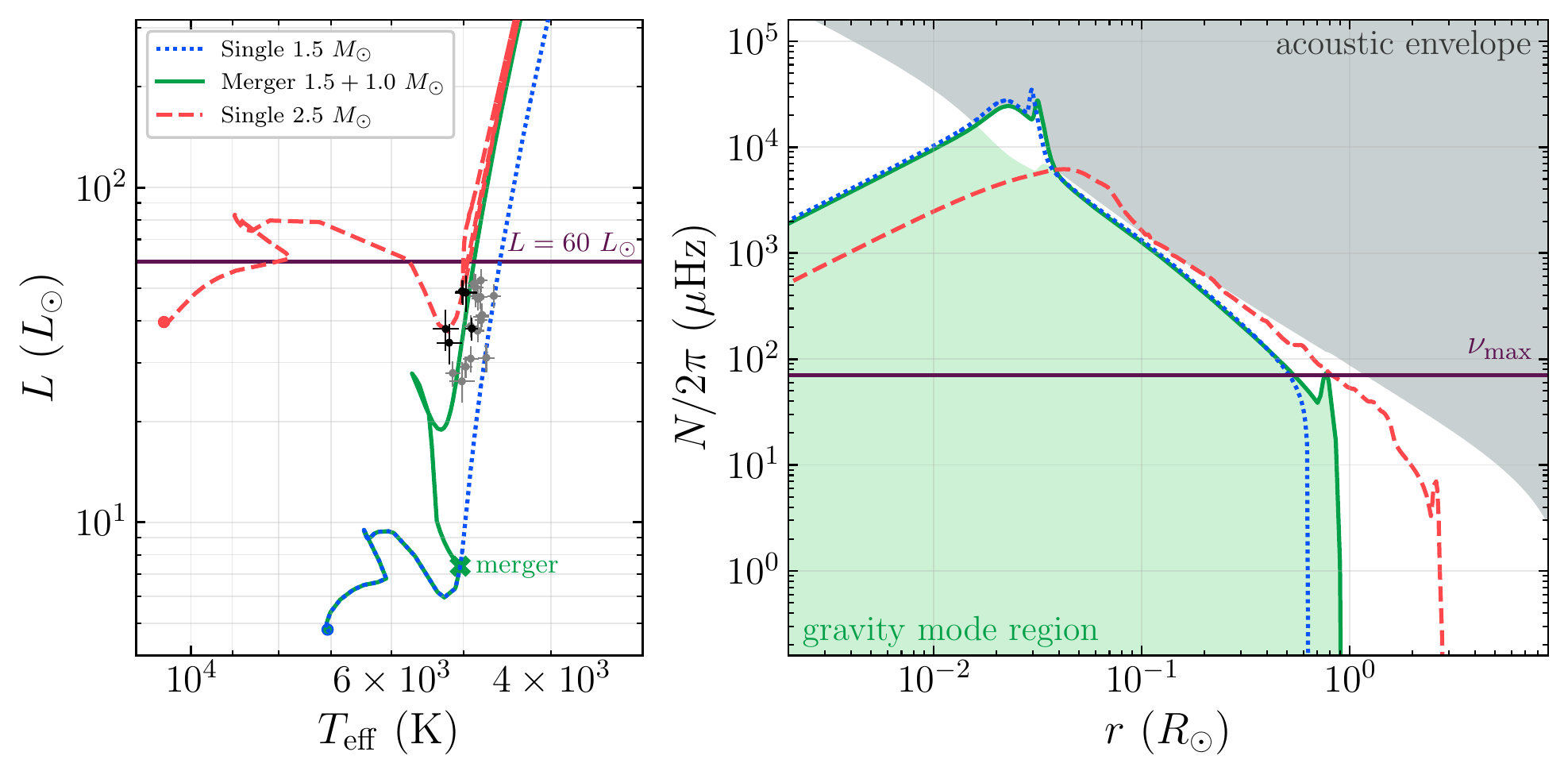}
    \caption{\textit{Left}: Hertzsprung-Russell diagram of a single $1.5$ $M_\odot$ star (\textit{blue dotted line}), a $1.5+1.0$ $M_\odot$ RG+MS merger product (\textit{solid green line}), and a single $2.5$ $M_\odot$ star (\textit{dashed red line}).
    Circular points indicate merger remnant candidates discussed in Section \ref{candidates}, with black points showing the best candidates.
    \textit{Right}: Propagation diagrams at $L=60$ $L_\odot$ for these three models.
    We also show the the gravity mode region (\textit{green}), as well as the acoustic mode region (\textit{gray}) and $\nu_{\mathrm{max}}$ (\textit{purple}), for the merger model, although the latter two are similar between all three models.
    Despite the merger, the Brunt--V\"ais\"al\"a frequency of the merger product most closely resembles that of a non-merged star of the original mass, rather than that of a non-merged star of the present-day mass.}
    \label{fig:three_models}
\end{figure*}

In this work, we demonstrate that RGs which have a merger in their histories can be identified via asteroseismology, provided that the merger occurs after the primary has left the main sequence (MS) and the secondary remains a MS star.
Asteroseismology will be effective at identifying a merger when the original RG develops a small, degenerate core that the final RG would not otherwise be expected to have.
The difference in core structure between such a merger product and an equal-mass RG forming via single star evolution manifests in a different gravity mode structure and, in turn, different period spacings of their dipole modes.
We sketch this picture heuristically in Figure \ref{fig:merger_cartoon}, which shows our fiducial comparison between a single $2.5$ $M_\odot$ RG and the product of a $1.5+1.0$ $M_\odot$ RG+MS merger.

\section{Stellar Models}

In order to obtain physically realistic stellar models, we employ Modules for Experiments in Stellar Astrophysics \citep[\textsc{mesa}, version r12778;][]{paxton2010modules,paxton2013modules,paxton2015modules,paxton2018modules,paxton2019modules}, an open-source one-dimensional stellar evolution code.
We first initialize a grid of single star models from $0.75$ $M_\odot$ from $2.75$ $M_\odot$, which are integrated through the MS and RGB, for the purpose of (1) providing initial conditions for binary merger models and (2) computing asteroseismic observables in single stars.
The stellar models are taken to be non-rotating and solar-metallicity, with reasonable values for convective overshoot. Model details and inlists are provided in Appendix \ref{mesa}.

We then model mergers as rapid accretion events with a rate $\dot{M}=10^{-5}$ $M_\odot$ yr$^{-1}$ at the surface of the star.
The original star starts off on the main sequence with solar composition, and at a specified age, it accretes material with the same composition as its surface (which is also close to solar composition).
While this cannot be expected to capture the transient structure of the star immediately after the merger, it should provide a reasonable model of the star after thermal relaxation, i.e., a few thermal times after merger.
While the adopted accretion rate is less than what is expected during a real merger, it should approximate a real merger event well because the accretion time scale $t_{\rm ac} = M/\dot{M} \sim10^5\,\mathrm{yr}$ is much shorter than a thermal time scale, hence the accretion is still in the rapid (adiabatic) regime.
We run a number of ``merger'' models, beginning with a fine grid of $1.5+1.0$ $M_\odot$ models where we vary the time of merger (Section \ref{time}). Next, we run a pair of grids where we vary the initial and final stellar masses, one in which the merger occurs when the primary is on the RGB, and the second when it is on the MS (Section \ref{massgrid}).
We then relax the convergence conditions required to run a $1.5+1.0$ $M_\odot$ model through helium burning, and we examine the behavior of the period spacing on the red clump (Section \ref{clump}).

While in reality the secondary star may be expected to penetrate deeply into the star's envelope before being disrupted and mixed into the star, modeling an RG+MS merger as a surface mass injection is sensible as long as the secondary mixes into the envelope before reaching the core.
We can use the approximation of \citet{eggleton1983approximations} to compute an effective Roche lobe radius for a secondary during a common envelope phase, taking as the mass ratio $q=M_2/M_{1,\mathrm{enc}}(r)$, the ratio of the mass of the secondary to the mass of the primary enclosed by the orbit.
For the $1.5+1.0$ $M_\odot$ merger shown in Figure \ref{fig:three_models}, we find that the MS secondary is expected to disrupt and mix into the primary's convective envelope at $r\approx2.8$ $R_\odot$ in our models, very close to the surface of the primary and far outside of the helium core ($r_{\mathrm{core}}\approx0.04$ $R_\odot$ on the lower RGB).

\subsection{Detailed oscillation mode calculations} \label{gyre}

While the period spacing between modes of the same gravity mode \strike{order}\revise{degree} generally lie quite close to the asymptotic period spacing $\Delta P_g$, they may deviate somewhat from this value, particularly when the mode has a large mixed character.
Since mixed modes are the most easily detected, it is critical to correct for this phenomenon when extracting the asymptotic period spacing from observations \citep[e.g.,][]{vrard2016period}.

In order to confirm that the asymptotic period spacing $\Delta P_g$ as defined by Equation \ref{eqn:gravity} lies close to the actual gravity mode spacing of our stellar profiles, we employ \textsc{gyre}, a shooting code which computes stellar oscillation modes given one-dimensional stellar profiles \citep{townsend2013gyre}.
Working in the adiabatic limit, we first calculate all oscillation modes with $\ell=0$ or $1$ lying within a factor of $2$ of $\nu_{\mathrm{max}}$ (computed from our \textsc{mesa} models using the scaling relation in Equation \ref{numax}) for single $1.5$ $M_\odot$ and $2.5$ $M_\odot$ RGs. We next compute these modes for the product of a $1.5+1.0$ $M_\odot$ merger that occurs soon after the main sequence when the primary's radius reaches $R=1.25R_{\mathrm{TAMS}}$. Here, $R_{\mathrm{TAMS}}$ is the radius of the star at the terminal age main sequence (TAMS), defined to be the earliest time that $X=0$ in the core.
Within each acoustic mode order, we calculate the difference in period between adjacent $\ell=1$ modes (which differ in gravity mode order), finding them to be very close to $\Delta P_g$ in almost all cases except when the character of the mode was very mixed.
Therefore, moving forwards, we center our discussion around $\Delta P_g$ (as defined in Equation \ref{eqn:gravity}), with the knowledge that (1) it is a good approximation to the actual, generally frequency-dependent period spacing, and (2) is typically reported in observations after accounting for this frequency dependence.

%
%
%

\section{Results}

\subsection{Heuristic description}

RGs comprise a large convective envelope surrounding a compact, high density core which primarily governs the star's evolution.
With the premise that a stellar merger between a RG with a degenerate core and a MS star mainly increases the RG's envelope mass while leaving its core intact, the core mass and structure of the RG is nearly unaffected by the merger.
Already compact and degenerate cores have higher Brunt--V\"ais\"al\"a frequencies in the core and thus a smaller $\Delta P_g$, relative to the less compact and less degenerate cores arising from more massive stars.
This gives rise to a robust observational signature of mergers of this type, provided that the core structure of the post-merger star is significantly different than that of a single RG star with the same mass as the merger product.

Figure \ref{fig:three_models} shows the Brunt--V\"ais\"al\"a frequency profiles of our fiducial models, where it is apparent that the RG merger product largely retains the core gravity mode structure of its RG progenitor.
Whereas the $1.5$ $M_\odot$ single star model has a more compact gravity mode region characterized by a larger value of $N$ in its degenerate core, the $2.5$ $M_\odot$ single star model has a more radially extended gravity mode region in its non-degenerate core whose $N$ peaks at a lower frequency.
Importantly, for an $1.5+1.0$ $M_\odot$ RG+MS merger, the Brunt--V\"ais\"al\"a frequency profile more closely resembles that of the original $1.5$ $M_\odot$ RG. 
As $\nu_{\mathrm{max}}$ and $\Delta\nu$ (together with $T_{\mathrm{eff}}$) provide an independent asteroseismic measurement of the mass, $\Delta P_g$ can be used to distinguish a merger product from a single star via their different gravity mode regions.
Specifically, merger remnants are expected to have a smaller $\Delta P_g$ (similar to that of the progenitor) relative to single stars of the same mass, when evaluated at the same luminosity.
In the model in Figure \ref{fig:three_models} at $L=60$ $L_\odot$, the merger remnant has $\Delta P_g=61.9\,\mathrm{s}$, very close to that of the original $1.5$ $M_\odot$ star ($\Delta P_g=57.7\,\mathrm{s}$) but very far from that of a single star $2.5$ $M_\odot$ ($\Delta P_g=164.0\,\mathrm{s}$).

The following sections elaborate on the point that $\Delta P_g$ reveals the fingerprint of a stellar merger, but only when the merger occurs after the primary has already left the MS (Section \ref{time}), and only when the merger brings an RG with a degenerate core into a mass regime where single star evolution does not produce degenerate cores (Section \ref{massgrid}).

\subsection{$\Delta P_g$ is sensitive to mergers on the RGB} \label{time}

\begin{figure*}
    \centering
    \includegraphics[width=\textwidth]{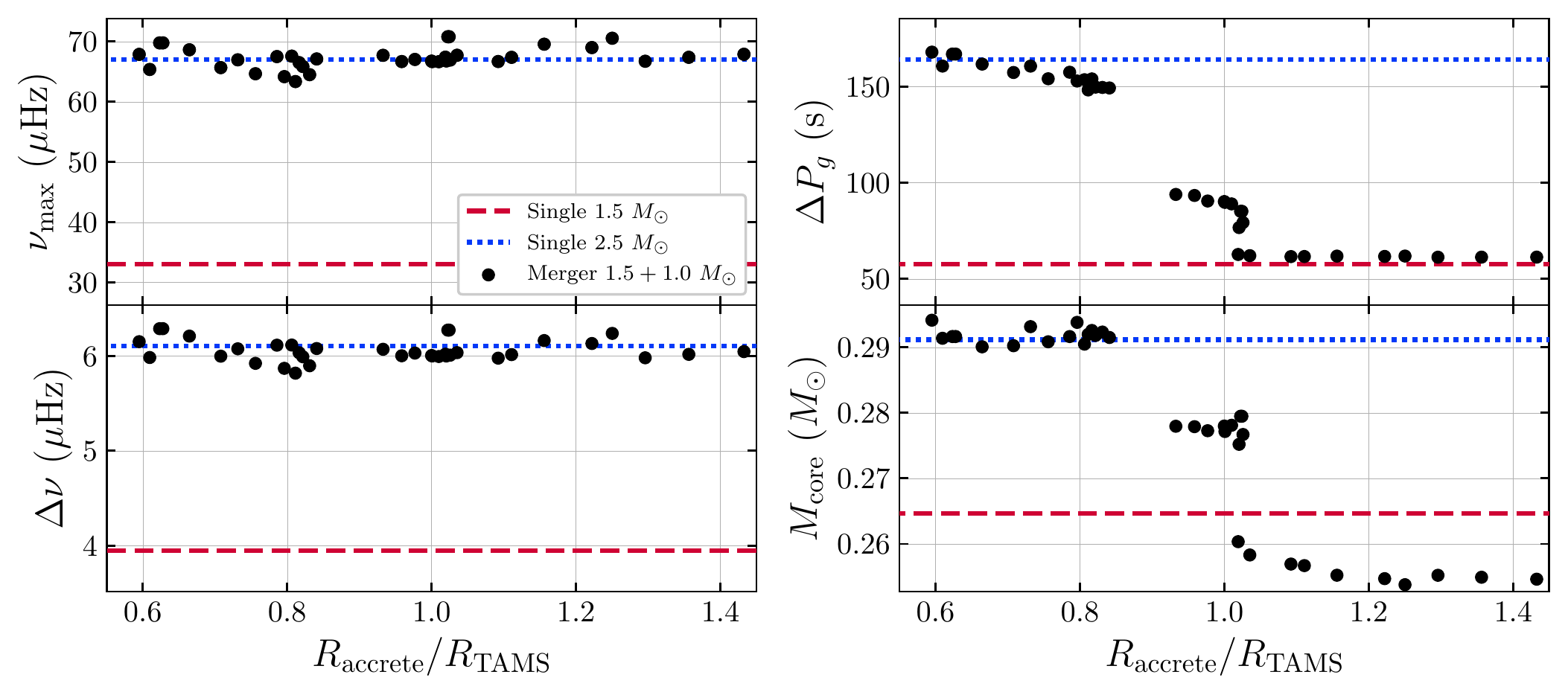}
    \caption{The frequency of maximum power $\nu_{\mathrm{max}}$ (\textit{top left}), large frequency spacing $\Delta\nu$ (\textit{bottom left}), asymptotic period spacing $\Delta P_g$ (\textit{top right}), and helium core mass (\textit{bottom right}) of the merger product of a $1.5$ $M_\odot$ primary with a $1.0$ $M_\odot$ secondary. The $x$-axis is the radius of the merging primary relative to its radius at the TAMS, so post-MS mergers occur when $R_{\rm accrete}/R_{\rm TAMS} > 1$. The asteroseismic quantities are evaluated when the merger remnant reaches $L=60$ $L_\odot$ on the RGB. 
    The red and blue lines show the analogous values for single $1.5$ $M_\odot$ and $2.5$ $M_\odot$ RG models of the same luminosity.}
    \label{fig:frequencies_with_radius}
\end{figure*}

We run a series of $1.5+1.0$ $M_\odot$ merger models where we vary the time of merger from the main sequence to the lower RGB.
When the merger product reaches $60$ $L_\odot$ during its ascent up the RGB, we calculate $\Delta P_g$---this luminosity is chosen because RGs at $L=60$ $L_\odot$ lie just below the bump at a value of $\Delta\nu$ where mixed modes are still observable, but well above the base of the RGB where the merger occurs.

We also calculate the large acoustic frequency spacing $\Delta\nu$ and \strike{central convective driving frequency}\revise{peak oscillation frequency} $\nu_{\mathrm{max}}$ \citep{brown1991detection}:
\begin{align}
    \Delta\nu &= \left(2\int^R_0\frac{dr}{c_s}\right)^{-1} \\
    \nu_{\mathrm{max}} &= 3100\,\mu\mathrm{Hz}\left(\frac{M}{M_\odot}\right)\left(\frac{R_\odot}{R}\right)^2\left(\frac{T_{\mathrm{eff},\odot}}{T_{\mathrm{eff}}}\right)^{1/2} \label{numax}
\end{align}
When combined with the surface temperature, these two quantities trace the total mass and radius of the star.
Independent of the time of merger, $\nu_{\mathrm{max}}$ and $\Delta\nu$ are unsurprisingly very close to their values for an equal-mass single RG, as the total mass and radius of the resulting RG will be almost identical to a non-merged analogy, when measured at the same luminosity.

Even though $\Delta\nu$ and $\nu_{\mathrm{max}}$ are only sensitive to the total mass and radius, $\Delta P_g$ traces the core structure and retains information about the star's evolutionary history which can be used to identify merger remnants.
Figure \ref{fig:frequencies_with_radius} demonstrates that for mergers occurring after the TAMS, $\Delta P_g$ of the merger product more closely resembles that of the $1.5$ $M_\odot$ progenitor as if it had never merged.
Physically, this is due to the pre-merger core already being high density and degenerate, such that its structure is insensitive to the overlying layers.
In other words, mergers which occur after the TAMS barely affect the underlying core structure of the progenitor.

\begin{figure*}
    \centering
    \includegraphics[width=\textwidth]{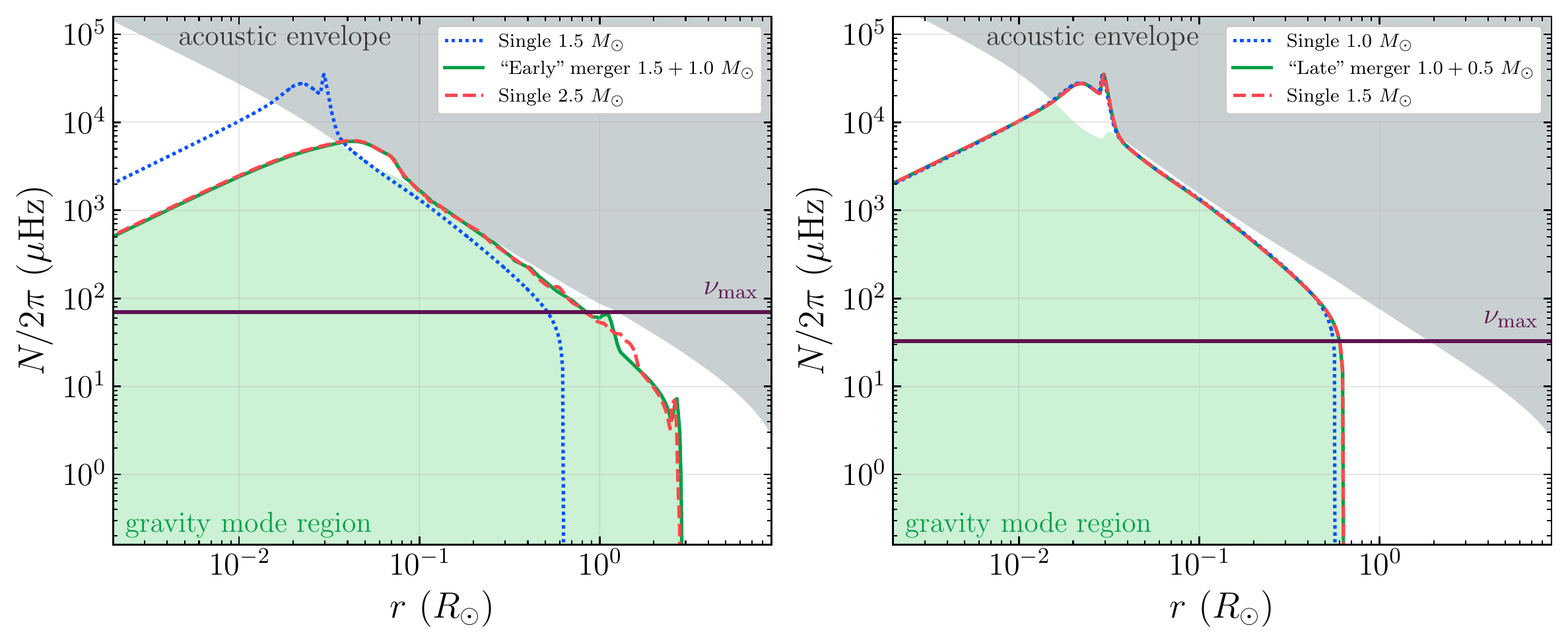}
    \caption{\textit{Left}: Propagation diagrams at $L=60$ $L_\odot$ for a single $1.5$ $M_\odot$ star (\textit{blue dotted line}), a $1.5+1.0$ $M_\odot$ MS+MS merger product (\textit{green solid line}), and a single $2.5$ $M_\odot$ star (\textit{red dashed line}).
    When the merger occurs while the primary is still on the MS,
    the merger product will be nearly indistinguishable via $\Delta P_g$ from an equal-mass single star.
    \textit{Right}: Propagation diagrams at $L=60$ $L_\odot$ for a single $1.0$ $M_\odot$ star (\textit{blue dotted line}), a $1.0+0.5$ $M_\odot$ RG+MS merger product (\textit{green solid line}), and a single $1.5$ $M_\odot$ star (\textit{red dashed line}).
    As single stars below $M\lesssim2$ $M_\odot$ all share similar degenerate core structures, $\Delta P_g$ cannot distinguish between a lower mass merger product and equal-mass single star.}
    \label{fig:propagation_models}
\end{figure*}

For mergers which occur when the primary is still on the MS, Figure \ref{fig:frequencies_with_radius} shows that $\Delta P_g$ is essentially the same as that for a single star of the same total mass.
The reason for this is revealed in Figure \ref{fig:propagation_models}, whose left-hand panel shows the propagation diagrams of an ``early'' $1.5+1.0$ $M_\odot$ merger which occurs at $t=0.5 \, t_{\mathrm{TAMS}}$, when the primary is on the MS.
We see that, in contrast to the ``late merger'' case (right panel, Figure \ref{fig:three_models}), the Brunt--V\"ais\"al\"a frequency profile of the ``early" merger model is indistinguishable from a single star of the same mass.
This occurs because the main sequence core is not degenerate and is sensitive to the mass of the overlying material, so the core readjusts to be nearly identical to that of a star that was born at $2.5$ $M_\odot$.
As an additional note, we find that, if the merger occurs very close to the TAMS, $\Delta P_g$ plateaus to an intermediate value between what is expected for $1.5$ $M_\odot$ and $2.5$ $M_\odot$ single stars (Figure \ref{fig:frequencies_with_radius}).
However, due to the short time window for this merger occur, it is unlikely that this case will be frequently observed.

In principle, a MS merger model may require more sophisticated simulations of the hydrodynamical mixing associated with such a traumatic event.
In such mergers, there may be a greater degree of mixing between the two stars, with material from both stars extending throughout the remnant in general.
However, as the evolution of a MS star of a given composition is essentially determined by its mass, we expect our simple surface accretion approximation to capture the most important effect.
To confirm this, we use \textsc{mesa}'s native entropy sorting procedure (accessible as \texttt{create\_merger\_model}) to model a $1.5+1.0$ MS+MS merger, and confirm that the resulting gravity mode structure at $L=60$ $L_\odot$ is virtually identical to the surface accretion merger model.



\subsection{Pre-merger core degeneracy is key to merger identification} \label{massgrid}

\begin{figure*}
    \centering
    \includegraphics[width=\textwidth]{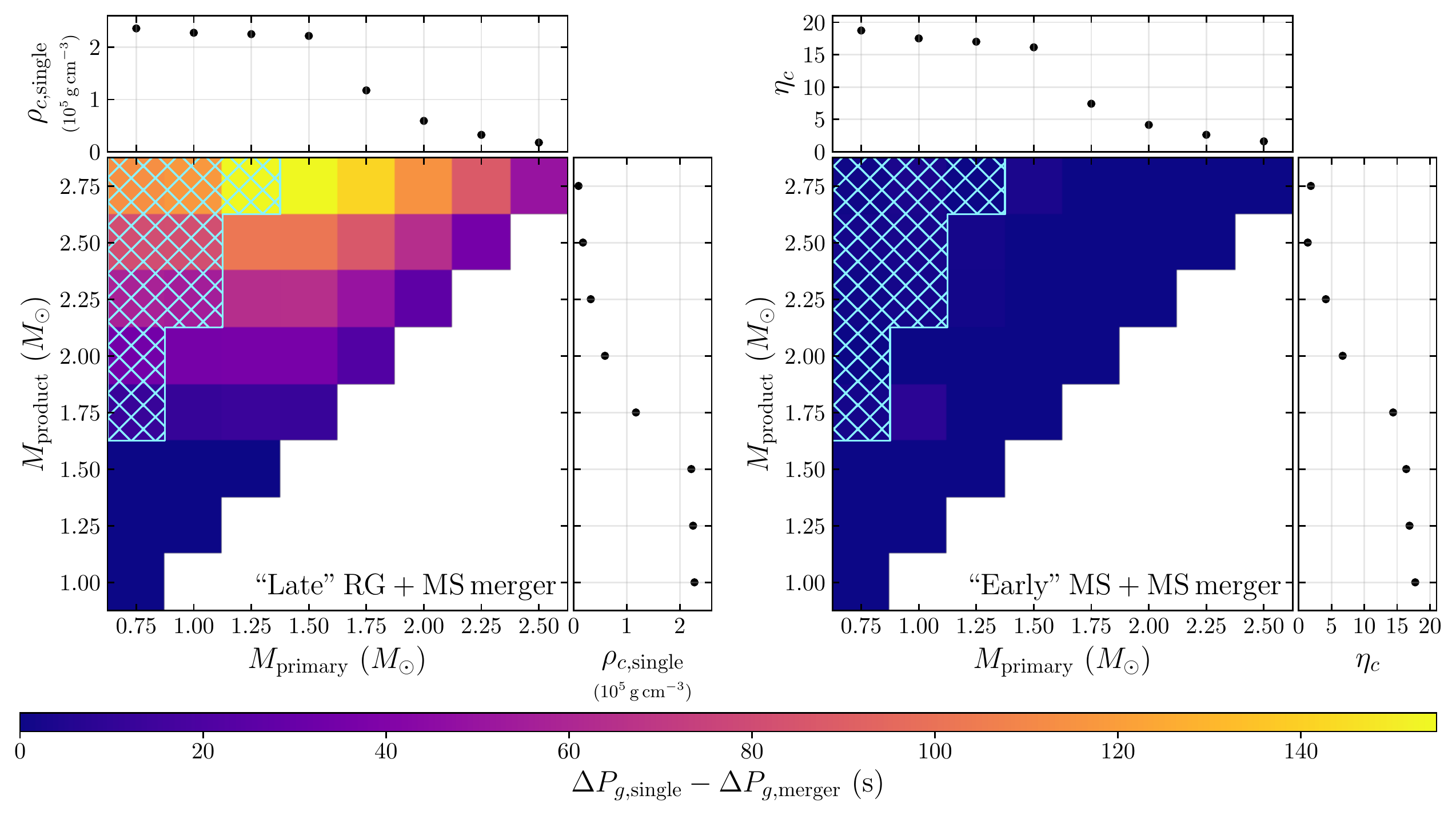}
    \caption{The difference between the period spacings of a single star and an RG+MS merger product of the same total mass, for a merger which occurs when the primary is on the lower RGB (\textit{left}) or on the MS (\textit{right}). In each panel, the period spacing difference is evaluated at a luminosity of $L=60$ $L_\odot$.
    The side panels show the central density $\rho_{c,\mathrm{single}}$ and central degeneracy parameter $\eta_c$ for single stars at $L=60$ $L_\odot$.
    Mergers resulting in stars with $M\gtrsim2$ $M_\odot$ are distinguishable from their non-merged counterparts via $\Delta P_g$, but only when the primary has already left the MS by the time the merger occurs.
    The hatching covers unphysical mergers where $M_{\mathrm{product}}>2M_{\mathrm{primary}}$.}
    \label{fig:merger_grid_2panel}
\end{figure*}

To understand the parameter space where merger products can be asteroseimically identified, we run two grids of merger models. The grids have primary mass in the range $M_{\mathrm{primary}}\in[0.75, 2.50]$ $M_\odot$ and post-merger mass in the range $M_{\mathrm{product}}\in[1.00, 2.75]$ $M_\odot$, spaced by $0.25$ $M_\odot$ in each dimension.
We consider both the case where the merger occurs on the lower RGB (when the primary's radius reaches $1.25$ times its value at TAMS) as well as the case where the merger occurs on the MS (when the primary reaches half of its TAMS age).
Figure \ref{fig:merger_grid_2panel} compares the period spacings of merger products at $60$ $L_\odot$ to non-merged stars of equal masses. It is clear that the period spacing is substantially different for high enough product masses when the merger occurs on the RGB, but it is practically indistinguishable when the merger occurs on the MS.
While $\Delta P_g$ could conceivably be reasonably discriminating for $M_{\mathrm{product}}$ as low as $\approx1.75$ $M_\odot$ in the RG+MS case, the effect is especially pronounced for $\gtrsim2$ $M_\odot$, corresponding approximately to the mass below which a star would be expected to develop a degenerate core.

In RGs with high core degeneracy, core properties such as the temperature, density, and Brunt--V\"ais\"al\"a frequency are primarily functions of the core mass, and they are largely independent of the properties of the surrounding envelope. This is the origin of the famous luminosity-core mass relation for RGs \citep{kippenhahn1981core}.
Hence, $\Delta P_g$ can be seen as a tracer for the core mass.
Late-stage MS stars will develop helium cores which grow as hydrogen-shell burning progresses, developing into proper RGs when the core mass reaches the star's Sch\"onberg--Chandrasekhar limit \citep{schonberg1942evolution}.
For a star with $M\lesssim2$ $M_\odot$, the core becomes degenerate before this limit is reached, and the star enters the RGB with a degenerate core \citep{cox1968principles}.
In this case, a merger which occurs after a degenerate core has already been formed will leave $\Delta P_g$ nearly unchanged---such mergers will simply add mass to the envelope, and the small increased pressure will leave the core unaffected.
The merger product will be distinguishable from a single star in the case that the latter would otherwise be expected to form a more massive non-degenerate core, which would have a larger $\Delta P_g$.

We note, however, that $\Delta P_g$ is insensitive to a merger in the case that a single star with the same mass as the product would have developed a degenerate core anyway---this can be seen in our models for $M_{\mathrm{product}}\lesssim2$ $M_\odot$ in Figure \ref{fig:merger_grid_2panel}.
This is also demonstrated in the right panel of Figure \ref{fig:propagation_models}, which shows a propagation diagram for the result of a $1.0+0.5$ RG+MS merger. Since both $1.0$ and $1.5$ $M_\odot$ single stars would be expected to develop degenerate cores through normal stellar evolution, their gravity mode structures are very similar when they evolve to the same point on the RGB, and $\Delta P_g$ cannot be used to distinguish them (or a merger bringing a $1.0$ $M_\odot$ RG to a $1.5$ $M_\odot$ RG).

\subsection{Mergers on the red clump are difficult to distinguish} \label{clump}

We have so far focused on first ascent giants, which manifest observationally as a roughly horizontal track at low $\Delta P_g$ tracing the star's evolution through $\Delta\nu$.
However, many\revise{,} more evolved RGs have \revise{already exhausted their hydrogen supplies available for off-center burning}\strike{already exhausted their core hydrogen supply} and have entered the helium core burning phase, with most such stars having accumulated on the red clump.
Red clump stars are also very apparent on a spacing diagram as a large cloud of points at low $\Delta\nu$ and high $\Delta P_g$, which makes them straightforwardly distinguishable from RGB stars (Figure \ref{fig:spacing_diagram}).
It is natural to wonder whether measurements of $\Delta P_g$ can be used to distinguish red clump merger remnants from single stars, similar to the process described above for RGB stars.

In practice, identifying merger remnants on the clump will be difficult due to the very similar values of $\Delta P_g$ between low-mass and high-mass clump stars.
The evolution of $\Delta P_g$ over time is similar for the single $1.5$ $M_\odot$ and merger $1.5+1.0$ $M_\odot$ models.
However, because the merger model has a larger total mass relative to the single star model, it has a systematically larger $\Delta\nu\simeq\sqrt{G\bar{\rho}}$.
This manifests as a slight horizontal offset between the two models' evolutionary tracks on a spacing diagram.
While this effect also applies to first ascent RGs, it is less obvious since the trajectory of such RGs through a spacing diagram is shallower, i.e., $\Delta\nu$ evolves much more quickly for first ascent giants than clump stars, relative to $\Delta P_g$.
This small offset between the single $1.5$ $M_\odot$ and merger $1.5+1.0$ $M_\odot$ evolutionary tracks is comparable to both models' offsets from the track of a single $2.5$ $M_\odot$ star.
In general, the three models all coincide with each other at some point in their evolution, making them difficult to distinguish using asteroseismology.
Therefore, although constraining the merger history using $\Delta P_g$ may be possible in some cases, we anticipate that it will be difficult for most clump stars.

\section{Merger Candidates}

\subsection{Promising candidates from $\Delta P_g$} \label{candidates}

\begin{figure*}
    \centering
    \includegraphics[width=\textwidth]{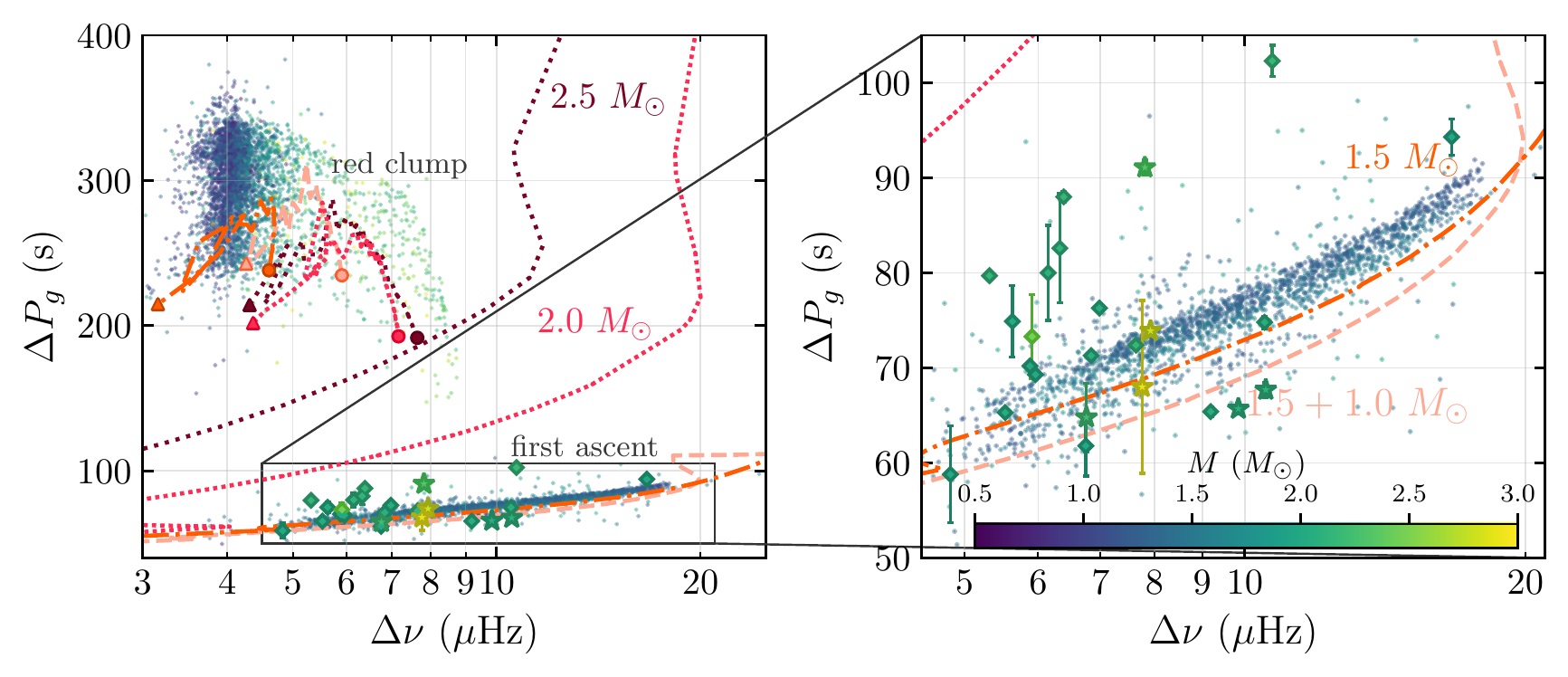}
    \caption{An asteroseismic frequency spacing diagram showing measurements of the g~mode period spacing $\Delta P_g$ as a function of the acoustic mode large frequency spacing $\Delta\nu$ from \textit{Kepler} data \citep{vrard2016period}.
    Points are color coded by their measured astroseismic mass, as shown by the color bar.
    Overlaid on this plot are evolutionary tracks along the RGB for single stars of mass $1.5 \, M_\odot$ (\textit{dash-dotted orange line}), $2.0\, M_\odot$ (\textit{pink dense-dotted}), and $2.5$ $M_\odot$ (\textit{maroon dotted}) stars, as well as a $1.5+1.0$ $M_\odot$ (\textit{salmon dashed}) merger product.
    The cloud of points in the upper left are clump stars, with model tracks covering a central helium fraction of $90\%$ (circle) to $1\%$ (triangle). The right panel zooms in on the RGB track, where stars evolve from right to left.
    Merger remnant candidates are emphasized, with star symbols indicating first ascent RGs with observed asteroseismic masses $M\geq2.0$ $M_\odot$ as reported by both \citet{vrard2016period} and \citet{yu2018asteroseismology}, and diamonds indicating those which have been reported to have $M\geq2.0$ $M_\odot$ by \citet{vrard2016period} only.
    These stars are inconsistent with single star models but lie near our merger tracks and are good candidates to be merger products.
    }
    \label{fig:spacing_diagram}
\end{figure*}

The evolutionary stage of RGs can be tracked on an asteroseismic period vs. frequency spacing diagram like that shown in Figure \ref{fig:spacing_diagram}, where merger remnants will appear as outliers relative to the paths taken by single stars. When ascending the RGB, RGs first evolve from larger $\Delta P_g$ and $\Delta\nu$ to smaller $\Delta P_g$ and $\Delta\nu$, later accumulating at high $\Delta P_g$ and low $\Delta\nu$ once they reach the red clump. Stars of different mass take different paths through the diagram, and merger remnants take different paths from single stars of the same mass. Hence, combined with an asteroseismic mass estimate (which can be deduced via $\nu_{\mathrm{max}}$, $\Delta\nu$, and $T_{\mathrm{eff}}$), mergers that occur after the primary has left the MS can readily be apparent from $\Delta P_g$. For stars with $M \gtrsim 2 \, M_\odot$, merger remnants will manifest as stars with a significantly lower $\Delta P_g$ than expected from their mass.
In other words, merger remnants will lie near tracks corresponding to lower mass single stars.

Figure \ref{fig:spacing_diagram} also shows the measured $\Delta P_g$ versus $\Delta\nu$ for a sample of \revise{$6111$}\strike{$5860$} RGs in the \textit{Kepler} field from the catalog of \cite{vrard2016period}\strike{, where we have excluded all stars which have been flagged as having aliases}.
Of these RGs, we coarsely classify these stars as first ascent giants (\revise{$1995$}\strike{$1745$}; $\Delta P_g<125 \, {\rm s}$) and red clump stars (\revise{$4116$}\strike{$4115$}; $\Delta P_g\geq125\, {\rm s}$).
We identify \revise{$24$}\strike{$15$} stars in this sample on the first ascent RGB with $M\geq2$ $M_\odot$ which, despite their ostensibly larger masses, appear to lie on the sequence of a less massive ($\approx \! 1.5 \, M_\odot$) star through this space.
\revise{Of these, $9$ are flagged as having aliases, 
although \citet{deheuvels2021seismic} find that the extracted period spacings of ``high-mass'' ($M\gtrsim1.6$ $M_\odot$) RGs within this sample are usually unaffected by these.}
Note that one of these stars (KIC 8517859) has been classified as a $\delta$ Scuti star in some catalogs \citet[e.g.,][]{barcelo2018envelope} which, if correct, would exclude it as a merger remnant candidate.
The possibility that it is a $\delta$~Scuti--RGB binary system \citep[similar to, e.g., the system reported by][]{murphy2021binary} is also intriguing.

However, \citet{yu2018asteroseismology}, who later revisited these asteroseismic mass measurements, found that \revise{$17$}\strike{$10$} of these candidates have masses $<2$ $M_\odot$ after applying a correction assuming that these stars are first ascent giants (with one candidate absent in their catalog).
Nonetheless, \revise{$6$}\strike{$4$} of these stars (KIC 12254159, KIC 2972876, \revise{KIC 7778197,} KIC 8708536, \revise{KIC 9907511,} and KIC 11465942) have $M>2$ $M_\odot$ in both catalogs, and should be considered the strongest merger remnant candidates in the sample.
These merger candidates are listed in Table \ref{tab:kic}.
\revise{For some of these candidates, it is possible that their true masses lie a few $\sigma$ below their reported values, in which case they are consistent with single stars with $M\simeq1.8$ $M_\odot$ without any need to invoke a merger scenario.}
\revise{Another possible source of false positives is that the extracted period spacings underestimate the true value---such errors may be exacerbated by suppressed dipole modes in some RGs \citep{mosser2012characterization,garcia2014droopy,fuller2015asteroseismology,stello2016suppression,mosser2017dipole}.}

Note that our $1.5$ $M_\odot$ single star model track appears to run at slightly lower $\Delta P_g$ than the observed sample. Lower mass models come closer to the data points due to their smaller frequency spacings, as expected since a mass of $\approx \! 1.2 \, M_\odot$ is most common amongst \textit{Kepler} RGs.
It may also be possible that the models predict slightly too small $\Delta P_g$ (or conversely, slightly too large $\Delta \nu$) or that this difference is related to a correction applied in the \citet{vrard2016period} sample in the conversion between the uncorrected period spacing (which depends complexly on coupling to acoustic modes) and the asymptotic value.


Because RG evolution is primarily governed by core physics, we expect that the product of a $1.5+1.0$ $M_\odot$ merger after the TAMS to ascend the RGB at a rate similar to a single $1.5 \, M_\odot$ star, i.e., much more slowly than a single $2.5$ $M_\odot$ star.
In our models, a $1.5+1.0$ $M_\odot$ merger remnant spends $37.8$ Myr in the range $4\,\mu\mathrm{Hz}\leq\Delta\nu\leq10\,\mu\mathrm{Hz}$ versus the much shorter $2.8$ Myr for a single $2.5$ $M_\odot$ star.
Hence, even though the single star evolutionary route is more common, merger products will be over-represented relative to single stars of the same mass within this range of frequency spacing.

\revise{Just before this paper was finalized, \citet{deheuvels2021seismic} performed a similar study investigating the asteroseismic signatures of mass transfer. They also found that RGs which lie below the main RGB sequence in $\Delta\nu$--$\Delta P_g$ space can be explained by stars that have accreted mass after developing a degenerate core. They propose that stars that lie below the main RGB sequence (especially those with $M \gtrsim 1.8 \, M_\odot$) have likely accreted mass, which increases their $\Delta\nu$ without modifying $\Delta P_g$ substantially. Indeed, our models predict the same behavior, which is why the merger model in Figure \ref{fig:spacing_diagram} lies to the right of the single star model.
Using this asteroseismic signature, they identify $\sim \! 30$ RGs which may have experienced mass transfer in the past.}
\revise{Several of their mass transfer candidates do not appear in our list of candidates because they either have $M<2 \, M_\odot$ or do not appear in the catalog of \citet{vrard2016period} but have separately measured period spacings.
In turn, our candidate list contains stars which do not lie below the RGB period spacing sequence and therefore are not selected by their method.}

\begin{table*}
    \centering
    \begin{tabular}{llllll}
        \hline
        Name & \multicolumn{2}{c}{$M$ ($M_\odot$)} & \multicolumn{2}{c}{$\Delta\nu$ ($\mu$Hz)} & \multicolumn{1}{c}{$\Delta P_g$ (s)} \\\cmidrule(lr){2-3}
        \cmidrule(lr){4-5}
        \cmidrule(lr){6-6}
        & \citet{vrard2016period} & \citet{yu2018asteroseismology} & \citet{vrard2016period} & \citet{yu2018asteroseismology} & \citet{vrard2016period} \\
        \hline
        KIC 2583386$^\star$ & $2.16\pm0.17$ & $1.92\pm0.12$ & $6.39$ & $6.339\pm0.032$ & $88.0\pm0.56$ \\
        KIC 2972876$^*$ & $2.33\pm0.25$ & $2.15\pm0.12$ & $7.81$ & $7.822\pm0.020$ & $91.1\pm0.81$ \\
        KIC 5341131$^{\ddag\star}$ & $2.04\pm0.10$ & $1.82\pm0.12$ & $5.63$ & $5.607\pm0.018$ & $74.9\pm3.77$ \\
        KIC 5385518$^\star$ & $2.15\pm0.15$ & $1.89\pm0.11$ & $5.32$ & $5.292\pm0.014$ & $79.7\pm0.37$ \\
        KIC 5820672$^\ddag$ & $2.08\pm0.10$ & $1.92\pm0.14$ & $10.50$ & $10.431\pm0.023$ & $74.8\pm0.64$ \\
        KIC 5857618$^\star$ & $2.13\pm0.14$ & $1.86\pm0.10$ & $6.33$ & $6.334\pm0.017$ & $82.6\pm5.73$ \\
        KIC 6118479 & $2.01\pm0.14$ & $1.60\pm0.09$ & $5.96$ & $5.938\pm0.019$ & $69.3\pm0.3$ \\
        KIC 6200178 & $2.06\pm0.16$ & $1.75\pm0.09$ & $6.98$ & $6.953\pm0.016$ & $76.3\pm0.46$ \\
        KIC 6437547 & $2.17\pm0.02$ & $1.90\pm0.10$ & $10.70$ & $10.669\pm0.023$ & $102.3\pm1.65$ \\
        KIC 7121674 & $2.02\pm0.08$ & $1.83\pm0.10$ & $5.88$ & $5.844\pm0.017$ & $70.2\pm0.31$ \\
        KIC 7457184 & $2.48\pm0.17$ & $1.96\pm0.12$ & $5.91$ & $5.913\pm0.017$ & $73.3\pm4.43$ \\
        KIC 7499531$^{\ddag\star}$ & $2.11\pm0.13$ & $1.99\pm0.10$ & $6.15$ & $6.115\pm0.014$ & $80.0\pm4.99$ \\
        KIC 7778197$^{*\ddag\star}$ & $2.09\pm0.12$ & $2.21\pm0.14$ & $9.84$ & $9.840\pm0.018$ & $65.7\pm0.55$ \\
        KIC 8055108$^\star$ & $2.07\pm0.13$ & $1.84\pm0.09$ & $9.19$ & $9.189\pm0.014$ & $65.4\pm0.50$ \\
        KIC 8277879$^\ddag$ & $2.16\pm0.09$ & $1.63\pm0.09$ & $7.64$ & $7.901\pm0.019$ & $72.4\pm0.48$ \\
        KIC 8364786 & $2.04\pm0.07$ & $1.79\pm0.10$ & $5.53$ & $5.510\pm0.013$ & $65.3\pm0.25$ \\
        KIC 8517859$^{\dagger\ddag}$ & $2.00\pm0.16$ & --- & $16.67$ & --- & $94.3\pm1.9$ \\
        KIC 8558329$^\ddag$ & $2.02\pm0.18$ & $1.65\pm0.10$ & $4.83$ & $4.821\pm0.013$ & $58.8\pm5.1$ \\
        KIC 8708536$^*$ & $2.82\pm0.15$ & $2.71\pm0.17$ & $7.92$ & $7.921\pm0.031$ & $73.9\pm0.57$ \\
        KIC 9329200 & $2.07\pm0.14$ & $1.54\pm0.09$ & $6.84$ & $6.823\pm0.016$ & $71.3\pm0.38$ \\
        KIC 9784586$^\star$ & $2.01\pm0.11$ & $1.73\pm0.09$ & $6.75$ & $6.728\pm0.012$ & $61.8\pm3.20$ \\
        KIC 9907511$^{*\ddag\star}$ & $2.12\pm0.19$ & $2.28\pm0.15$ & $10.53$ & $10.488\pm0.02$ & $67.7\pm0.80$ \\
        KIC 11465942$^*$ & $2.92\pm0.14$ & $2.62\pm0.21$ & $7.76$ & $7.762\pm0.026$ & $68.0\pm9.1$ \\
        KIC 12254159$^*$ & $2.23\pm0.11$ & $2.12\pm0.13$ & $6.76$ & $6.732\pm0.012$ & $64.8\pm3.6$ \\
        \hline
    \end{tabular}
    \caption{Candidate merger products from Figure \ref{fig:spacing_diagram}, identified as stars with period spacings consistent with low-mass RGB stars but with asteroseismic masses $\geq2.0$ $M_\odot$ (Section \ref{candidates}). Asteroseismic properties are taken from the catalog of \citet{vrard2016period} and \citet{yu2018asteroseismology}---for the latter, we have reported the corrected mass where the star has been assumed to be a first ascent RG.\\
    ${}^*$ Stars where \citet{yu2018asteroseismology} also find $M\geq2.00$ $M_\odot$. 
    These stars should be taken as the strongest merger remnant candidates.\\
    ${}^\dagger$ KIC 8517859 does not appear in the catalog of \citet{yu2018asteroseismology}.
    In addition, it has been classified as a $\delta$ Scuti variable in some catalogs, which would exclude it as a viable merger remnant candidate if true.\\
    ${}^\ddag$ Stars which are not present in the \citet{gaulme2020active} catalog, which examined stars for surface activity.\\
    ${}^\star$ Stars whose observations are flagged as having aliases in the catalog of \citet{vrard2016period}.
    }
    \label{tab:kic}
\end{table*}

\subsection{Recognizing merger remnants in the absence of $\Delta P_g$} \label{sec:nopg}

While $\Delta P_g$ is a robust way to identify certain merger remnants on the basis of an apparently under-massive core, it may be possible to identify merger remnants without $\Delta P_g$.
Specifically, by using measurements of $\Delta\nu$, $\nu_{\mathrm{max}}$, and $T_{\mathrm{eff}}$, one can in principle constrain the stellar mass $M$, radius $R$, and luminosity $L$.
The latter quantities can also be determined using \textit{Gaia} parallax measurements, given a reliable $T_{\rm eff}$.
A merger remnant would then manifest as a giant which is less luminous and/or cooler than possible for a single star of mass $M$ which is just beginning its ascent up the RGB.

For example, one could distinguish a $1.5+1.0$ $M_\odot$ merger remnant from a single $2.5$ $M_\odot$ star based on their location on the Hertzsprung-Russell diagram (HRD) in Figure \ref{fig:three_models}.
This is likely only possible if the remnant giant is young enough to be located at the base of the RGB such that its luminosity is smaller than that of a single $2.5$ $M_\odot$ model at the bottom of the RGB.
In other words, at a given $T_{\rm eff}$ and mass $M$, merger remnants would be expected to have smaller luminosity (i.e., smaller $R$ and larger $\Delta \nu$ and $\nu_{\rm max}$) than expected to be possible from the model track of a single star.
As an example, four of our merger candidates from above lie below the $2.5 \, M_\odot$ track in Figure \ref{fig:three_models} and could potentially be identified using this method, if they were to have asteroseismic masses greater than $2.5 \, M_\odot$.

This method of identification is limited because it can only identify remnants young enough that they lie near the base of the RGB, and with masses where the minimum RGB luminosity is somewhat sensitive to mass.
Additionally, the stellar track on a HRD is model-dependent and can vary with metallicity, further complicating this method \citep{basu2012effect}.
Nonetheless, this method does not require measurements of $\Delta P_g$, so it may be applicable to a much larger number of stars for which only $\Delta \nu$ and $\nu_{\rm max}$ can be measured, as expected for the bulk of red giants observed by \textit{TESS}.
We encourage follow-up work to investigate this technique in more detail.

\section{Discussion}

\subsection{Merger dynamics} 

The RG+MS mergers described in this work may naturally be formed by binary coalescences when the primary in a close binary expands along the RGB and initiates unstable mass transfer \citep[for a review, see][]{ivanova2013common}.
For conservative mass transfer, a mass ratio $q<2/3$ (where $q$ is the ratio of the donor to accretor mass) is required for stable mass transfer from a $n=3/2$ polytrope. Hence mass transfer in standard coeval binaries (where an RG primary accretes onto a less massive secondary such that $q > 1$) is typically expected to be unstable, though we note a radiative core does enhance mass transfer stability \citep{soberman1997stability}. In these unstable cases, stars are expected to eventually merge in a bright transient \citep[``luminous red nova''; e.g.,][]{ivanova2013identification}. Moreover, on the lower RGB where the envelope binding energy is still large, mergers will occur more frequently relative to successful envelope ejections.

Hydrodynamical simulations \citep{macleod2018bound} have shown that stellar coalescences are expected to produce a bipolar outflow structure which has been observed in follow-up radio observations of a number of luminous red novae \citep{kaminski2018submillimeter}.
With observations taken using the Atacama Large Millimeter Array, \citet{kaminski2018submillimeter} estimate the ejecta mass of three red novae (V4332 Sgr, V1309 Sco, and V838 Mon) as varying dramatically between events, but characteristically on the level of a few percent of the total mass of the system.
This is comparable to the prediction of \cite{metzger2017merger} that the ejecta mass $M_{\mathrm{ej}}\sim0.1M$ is a relatively small fraction of the total mass of the system, such that the merger results in a single star with nearly the same total mass.
The detection of similar outflow material around stars identified as merger products using asteroseismology could validate this method.
In may cases, however, the merger ejecta may have already been expelled from the system.
The lifetime of protoplanetary disks has been estimated to be on the order of 1 Myr \citep{mamajek2009initial,cieza2015structure,li2016lifetimes}, whereas $\Delta P_g$ should be able to discern a merger remnant for approximately $\approx40$ Myr throughout its ascent up the lower RGB. Hence, circumstellar disks have most likely already been expelled from most asteroseismically detectable merger remnants.

Interestingly, V1309 Sco \citep{tylenda2011V1309Sco} is thought to have arisen from a merger involving a primary of mass $M_1\sim \! 1.5 \, M_\odot$ and radius $R_1 \sim \! 3.5 \, R_\odot$ \citep{metzger2017merger}. The primary was thus on the sub-giant branch at the time of merger, at a time favorable for asteroseismic identification of the merger product. For a secondary of mass $M_2 \gtrsim 0.25 \, M_\odot$, the merger product would lie in the mass range favorable for asteroseismic identification, so V1309 Sco may be a perfect example of the type of stellar merger whose remnant can later be identified through asteroseismic techniques.
Along similar lines, the \textit{SPIRITS} survey recently identified a class of ``eSPecially Red Intermediate-luminosity Transient Events'' \citep[``SPRITEs'';][]{kasliwal2017spirits}, characterized by luminosities between those of novae and supernovae, relatively red colors, and lack of any optical counterparts.
\citep{metzger2017merger} suggest that these dustier SPRITE events may in fact be giant star mergers, in contrast to luminous red novae, which are more likely to be MS mergers.

\subsection{Additional merger signals}
\label{othersigns}

To corroborate the merger candidates asteroseismically identified above, additional evidence for a previous merger event would be useful. Merger remnants are expected to initially be rapidly rotating, though they may spin down rapidly on a time scale of less than 1 Myr \citep{casey2019lirich}.
Some remnants may be expected to exhibit large magnetic fields generated during the merger \citep{schneider2016rejuvenation,schneider2019stellar} or sustained by a dynamo in the convective envelope due to the high post-merger rotation rate.
The fields may be detected via spectropolarimetry \citep{auriere2015spectropolarimetry} or they may manifest in Ca II H\&K emission \citep{deMedeiros1999catalog} or X-ray emission \citep{soker2007magnetic}.
A class of lithium-enriched giants has also emerged in the last few decades, some of which are also rapidly rotating \citep{charbonnel2000nature,drake2002rapidly,rebull2015infrared,martell2020galah}.
Evidence has suggested tidal spin-up \citep{casey2019lirich}, stellar mergers \citep{siess1999accretion,jura2003flared,melis2020rise} and/or giant planet accretion \citep{,denissenkov2000episodic,sandquist2002critical,reddy2002spectroscopic,carlberg2012observable,punzi2017young,soares2020lithium} as explanations for these lithium-enhanced, sometimes rapidly rotating stars.
\strike{Naturally, such objects are promising candidates for asteroseismic follow-up.}
\revise{Asteroseismic merger candidates should be examined for these other signatures of a prior stellar merger.}

While there have been many surveys that have searched for lithium enhancement in RGs, many of them have were directed at a different field of view than the \textit{Kepler} data set \citep{buder2018galah,kumar2020discovery}, not sufficiently photometrically deep \citep[e.g.,][]{kumar2011origin}, or restricted only to clump stars \citep[e.g.,][]{singh2021tracking}.
The studies of \cite{deepak2021lithium} and \cite{yan2021most} cross-referenced asteroseismic classifications with high lithium abundances via LAMOST data, but none of our merger candidates appear in their \revise{publicly available} samples, suggesting they are likely not strongly lithium enhanced.
\revise{While the catalog of \citet{casey2019lirich} contains $23$ lithium-rich giants which have also been asteroseismically observed by \textit{Kepler}, only $2$ of them have been identified to be on the RGB, and none of them coincide with any of our $24$ merger remnant candidates.}
Additional spectroscopic study of our candidates may reveal more subtle, unusual compositional features which may be associated with a previous merger.

In addition, examining the light curves of \textit{Kepler} RGs (and performing limited spectroscopic follow-up), \citet{gaulme2020active} find a correspondence between surface activity, close binarity, and suppressed oscillations, consistent with previous work \citep{garcia2010corot,chaplin2011evidence,gaulme2014surface,mathur2019revisiting}.
As discussed in Section \ref{othersigns}, merger remnants may have elevated rotational rates and magnetic fields, suggesting that their oscillations may be preferentially suppressed.
This may prevent a measurement of $\Delta P_g$ in some cases and may partially account for our relatively low fraction of remnant candidates (see Section \ref{rates}). 
Of our \revise{$24$}\strike{$15$} remnant candidates, \revise{$16$}\strike{$11$} appear within the catalog of \citet{gaulme2020active}, who search for surface activity via rotational modulation in RG's light curves.
However, they do not report surface activity in any of these candidates.
In addition, none of our candidates appear in the rotational catalog of
\citet{ceillier2017surface}.
While these non-detections do not provide additional support for the merger hypothesis, \revise{they}\strike{it} may reflect an asteroseismic candidate selection bias due to the suppression of oscillations associated with stronger magnetic activity.

\subsection{Rates of stellar mergers}
\label{rates}

For a circular orbit and mass ratio $q=3/2$, Roche overflow will occur when $a=2.4R_1$ \citep{eggleton1983approximations}.
We calculate that a $1.5$ $M_\odot$ in a circular binary with a $1.0$ $M_\odot$ star will undergo Roche overflow on the lower RGB ($a\lesssim30$ $R_\odot$) when the period $P\lesssim12$ d, with weak dependence on the mass ratio.
Such close binaries should account for $\approx4\%$ of solar-type binaries \citep{raghavan2010survey}.
\citet{price2020close} demonstrate a deficit of ``close’’ binaries in red clump and asymptotic giant stars suggestive of stellar mergers on the RGB.
Their observed decrease of close binaries approaching the red clump implies that $\approx8\%$ of systems (singles and binaries) merge on the RGB, with $\approx3\%$ of stars merging on the lower RGB where $\log g$ is higher than that of the clump but lower than that of the MS.
Tracking transient events, \citet{kochanek2014stellar} additionally find that the rate of mergers in the Milky Way between an MS star and an evolved star is $\approx0.045$ yr$^{-1}$.
Together with their star formation model ($3.5$ $M_\odot$ yr$^{-1}$ and the initial mass function of \citealt{kroupa2003galactic}) this merger rate implies that $\approx \! 7\%$ of red giants are merger remnants (although many of these mergers may occur higher up on the RGB).
These observations consistently suggest that the fraction of lower RGB stars that are merger remnants (and which merged after the MS) is on the order of a few percent.

Within the \citet{vrard2016period} data set, we identify \revise{24}\strike{$15$} candidate remnants (see Section \ref{candidates}), representing \revise{$\approx \! 1.2\%$}\strike{$\approx \! 0.9\%$} of the total number of RGB stars in their sample.
A total of \revise{$6$ ($\approx \! 0.3\%$)}\strike{$4$ ($\approx \! 0.2\%$)} are found to be strong candidates using the asterosemismic mass measurements of both \citet{vrard2016period} and \citet{yu2018asteroseismology}.
These fractions appear to fall somewhat short of our estimates above, but this is not unexpected.
Our method is most sensitive to the subset of RG remnants with $M\gtrsim2$ $M_\odot$ (corresponding to the identification criteria for candidate remnants), and also those which merge low enough on the RGB to produce a remnant which can still be probed effectively by asteroseismology. More detailed population synthesis would be needed to confirm whether our candidate fraction of $\sim \! 1\%$ is consistent with expectations of merger rates fulfilling the asteroseismic selection criteria.

\subsection{Mergers in dense stellar environments}

Dense stellar populations are clearly a natural setting for frequent stellar collisions as well as stellar-evolution mediated mergers in binaries hardened by scattering events.
The high stellar densities associated with the core of globular clusters make them hotbeds for such mergers. \citet{hills1976stellar} estimate that as many as tens of percent of stars in some globular clusters may have suffered from at least one collision in their history, and \citet{liu2021fractions} find (assuming an initial binary fraction $f_b=0.5$) that as many as 50\% of RGs in a globular cluster may have undergone a binary interaction, with evolved blue straggler stars making up $\simeq10\%$ of RGs.
Unfortunately, owing to limited observing fields and stellar crowding, asteroseismic measurements of stars in star clusters is sparse---only four open clusters appear in the \textit{Kepler} field, and only two of those have measured period spacings for non-clump giants \citep[NGC 6791 and NGC 6819;][]{corsaro2012asteroseismology}.
\revise{Using these data,  \citet{brogaard2021asteroseismology} recently demonstrated the presence of overmassive giants in NGC 6791---these stars likely originate from mass transfer events or mergers which could potentially be encoded in the period spacing.}
In addition, while limited asteroseismology has been conducted on globular clusters  \citep[e.g.,][]{stello2009solar,miglio2016detection}, measurements of the period spacing for stars in these clusters still remain elusive.


Ultimately, future observational asteroseismology campaigns, especially those directed towards dense stellar regions \citep[e.g.,][]{miglio2019haydn}, appear lucrative for identifying a large sample of merger products as well as providing a direct measurement of the merger rate in these populations.
Optimal observing targets for this type of merger remnant identification would have turn-off masses $\sim1$--$2$ $M_\odot$, where stars with masses $\lesssim2$ $M_\odot$ have entered the RGB but $M\gtrsim2$ $M_\odot$ merger remnants can still form (although this threshold mass may decrease somewhat at lower metallicity).
Such populations would place detectable merger remnants below the RG bump, where asteroseismology is most effective.

\section{Conclusion}

In this work, we investigated the asteroseismic signatures of stellar mergers, focusing on observable diagnostics in red giant merger remnants. Our main finding is that merger remnants can often be identified by the presence of an over-massive envelope relative to their cores, compared to what is expected for a single star. 
Merger remnants can be found amongst red giants, provided an asteroseismic measurement of the mass (via $\nu_{\mathrm{max}}$, $\Delta\nu$, and $T_{\mathrm{eff}}$), in addition to a measurement of the mixed mode period spacing $\Delta P_g$. Since the latter traces the core structure, it can be used to distinguish merger products from single stars under the following conditions:
\begin{itemize}
    \item The merger occurs when the primary is on the RGB, so that it has already developed a dense core and the merger essentially only adds to the envelope of the star (Section \ref{time}).
    \item The additional mass contributed by the secondary brings the mass of the giant from $M\lesssim2$ $M_\odot$ to $M\gtrsim 2$ $M_\odot$.
    This threshold corresponds roughly to the mass below which an RG would form a degenerate core, which would be distinguishable from the non-degenerate core of a more massive star formed via single star evolution (Section \ref{massgrid}).
\end{itemize}
Mergers that occur when the primary is on the main sequence are difficult to identify because the merger remnant structure is nearly indistinguishable from a single star of the same mass. The same is true for a merger that does not bring the total mass above $\approx \! 2 \, M_\odot$.

In other words, $\Delta P_g$ can be used to identify a merger remnant in the situation where the primary in a RG+MS merger has already developed a dense and degenerate core that withstands the merger, which would not otherwise be produced by a single star with the mass of the merger product. At the same point on the HRD (or alternatively, at the same $\nu_{\rm max}$ or $\Delta \nu$), a merger product is distinguished by a smaller period spacing relative to the expectations of a single star (Figure \ref{fig:spacing_diagram}).
Even without a $\Delta P_g$ measurement, mergers remnants can also potentially be identified as stars having a luminosity that is too low for their asteroseismically measured mass (see Section \ref{sec:nopg}), and future work should examine this possibility in more detail.

Fortunately, the RG mass range where merger remnants can be identified is well-sampled in existing asteroseismic catalogs built primarily from \textit{Kepler} data.
Using the catalog of \cite{vrard2016period}, we have identified \revise{$24$}\strike{$15$} promising candidates in Section \ref{candidates}, and we encourage follow-up observations to search for additional hints of a prior merger such as rapid rotation, magnetic fields, unusual chemical abundances, or circumstellar gas and dust.
These stars are a natural endpoint of close binary stellar evolution, and they are expected to be even more common in dense stellar environments.
A further examination of the data and future observational surveys will provide illuminating constraints on the occurrence rates and outcomes of stellar mergers in the Milky Way.

\section*{Acknowledgements}

This work was inspired by discussions at the 2019 Scialog Meeting. We thank Dan Huber, Carl Melis, \revise{S\'ebastien Deheuvels,} Fred Rasio, Jamie Lombardi, Tuguldur Sukhbold, and Pablo Marchant for insightful discussions and advice both scientific and technical.
N.Z.R. acknowledges support from the Dominic Orr Graduate Fellowship at Caltech.
J.F. is thankful for support through an Innovator Grant from The Rose Hills Foundation, and the Sloan Foundation through grant FG-2018-10515.
\revise{This research has made use of the SIMBAD database,
operated at CDS, Strasbourg, France.}
\revise{We thank the anonymous referee for their useful suggestions.}

\section*{Data Availability}

Inlists and selected output files for the \textsc{mesa} simulations described in this work are available in an accompanying Zenodo repository\footnote{\href{https://doi.org/10.5281/zenodo.4782723}{https://doi.org/10.5281/zenodo.4782723}} \citep{nicholas_rui_2021_4782723}.

\bibliographystyle{mnras}
\bibliography{bibliography}

\appendix

\definecolor{codegreen}{rgb}{0,0.6,0}
\definecolor{codegray}{rgb}{0.5,0.5,0.5}
\definecolor{codepurple}{rgb}{0.58,0,0.82}
\definecolor{backcolour}{rgb}{0.95,0.95,0.92}

\lstdefinestyle{mystyle}{
    backgroundcolor=\color{backcolour},   
    commentstyle=\color{codegreen},
    keywordstyle=\color{magenta},
    numberstyle=\tiny\color{codegray},
    stringstyle=\color{codepurple},
    basicstyle=\ttfamily\footnotesize,
    breakatwhitespace=false,         
    breaklines=true,                 
    captionpos=b,                    
    keepspaces=true,                 
    numbers=left,                    
    numbersep=5pt,                  
    showspaces=false,                
    showstringspaces=false,
    showtabs=false,                  
    tabsize=2
}

\lstset{style=mystyle}

\section{MESA Simulation Controls}
\label{mesa}

The input \texttt{inlist\_project} files for all of our \textsc{mesa} runs are very similar, with variations in specific parameters which control the initial mass of the star, as well as the mass accreted during the rapid merger period.
Here, the parameters \texttt{x\_ctrl(1)} and \texttt{x\_ctrl(2)} represent the mass accretion rate during the merger period (fixed at $\dot{M}=10^{-5}$ $M_\odot$ yr$^{-1}$) and the star age at which the merger occurs, respectively.
We have also used the parameter \texttt{x\_integer\_ctrl(1)} to control profile write-out.
This work is accompanied by a Zenodo repository containing inlists and selected output files associated with the simulations used in this work \citep{nicholas_rui_2021_4782723}.
We have included as a representative example the \texttt{inlist\_project} file for the $1.5+1.0$ $M_\odot$ merger run, where the merger occurs when the radius of the primary reaches $125\%$ of its value at TAMS:

\begin{lstlisting}[language=Fortran]
&star_job
    pgstar_flag = .true.
/ ! end of star_job namelist

&controls
    !---------------------------------------- Write GYRE
    write_pulse_data_with_profile = .true.
    pulse_data_format = 'GYRE'

    x_integer_ctrl(1) = 10 ! Force write-out at log L close to integer values divided by this number

    !---------------------------------------- Manages accretion
    x_ctrl(1) = 1e-5 ! mass accretion rate
    x_ctrl(2) = 2614839409.7825627 ! time (yr) at which to start accretion (if 0, no accretion)

    mass_change = 0 ! initial accretion rate (modified dynamically)
    max_star_mass_for_gain = 2.50

    !----------------------------------------  MAIN
    initial_mass = 1.50
    initial_z = 0.02
    use_Type2_opacities = .true.
    Zbase = 2.d-2

    predictive_mix(1) = .true.
    predictive_superad_thresh(1) = 0.005
    predictive_avoid_reversal(1) = 'he4'
    predictive_zone_type(1) = 'any'
    predictive_zone_loc(1) = 'core'
    predictive_bdy_loc(1) = 'top'

    dX_div_X_limit_min_X = 3d-5
    dX_div_X_limit = 3d-1
    dX_nuc_drop_min_X_limit = 3d-5
    dX_nuc_drop_limit = 3d-2

    !----------------------------------------  WIND
    cool_wind_RGB_scheme = 'Reimers'
    cool_wind_AGB_scheme = 'Blocker'
    RGB_to_AGB_wind_switch = 1d-4
    Reimers_scaling_factor = 0.2
    Blocker_scaling_factor = 0.5
    
    use_accreted_material_j = .true.
    accreted_material_j = 0
    
    !----------------------------------------  OVERSHOOTING
    overshoot_scheme(1) = 'exponential'
    overshoot_zone_type(1) = 'nonburn'
    overshoot_zone_loc(1) = 'core'
    overshoot_bdy_loc(1) = 'top'
    overshoot_f(1) = 0.015
    overshoot_f0(1) = 0.005
    
    overshoot_scheme(2) = 'exponential'
    overshoot_zone_type(2) = 'nonburn'
    overshoot_zone_loc(2) = 'shell'
    overshoot_bdy_loc(2) = 'any'
    overshoot_f(2) = 0.015
    overshoot_f0(2) = 0.005
    
    overshoot_scheme(3) = 'exponential'
    overshoot_zone_type(3) = 'burn_H'
    overshoot_zone_loc(3) = 'core'
    overshoot_bdy_loc(3) = 'top'
    overshoot_f(3) = 0.015
    overshoot_f0(3) = 0.005
    
    overshoot_scheme(4) = 'exponential'
    overshoot_zone_type(4) = 'burn_H'
    overshoot_zone_loc(4) = 'shell'
    overshoot_bdy_loc(4) = 'any'
    overshoot_f(4) = 0.015
    overshoot_f0(4) = 0.005
    
    overshoot_scheme(5) = 'exponential'
    overshoot_zone_type(5) = 'burn_He'    
    overshoot_zone_loc(5) = 'core'
    overshoot_bdy_loc(5) = 'top'
    overshoot_f(5) = 0.015
    overshoot_f0(5) = 0.005
    
    overshoot_scheme(6) = 'exponential' 
    overshoot_zone_type(6) = 'burn_He'
    overshoot_zone_loc(6) = 'shell'
    overshoot_bdy_loc(6) = 'any'
    overshoot_f(6) = 0.015
    overshoot_f0(6) = 0.005
    
    set_min_D_mix = .true.
    min_D_mix = 1d0

    !----------------------------------------  MISC
    photo_interval = 25
    profile_interval = 50
    max_num_profile_models = 3000
    history_interval = 10
    terminal_interval = 10
    write_header_frequency = 10
    max_number_backups = 500
    max_number_retries = 3000
    max_timestep = 3d15
    
    !----------------------------------------  MISC
    photo_interval = 25
    profile_interval = 50
    max_num_profile_models = 3000
    history_interval = 10
    terminal_interval = 10
    write_header_frequency = 10
    max_number_backups = 500
    max_number_retries = 3000
    max_timestep = 3d15

    !----------------------------------------  MESH
    mesh_delta_coeff = 1
    varcontrol_target = 0.7d-3
    
/ ! end of controls namelist

&pgstar
\end{lstlisting}

The \texttt{run\_star\_extras.f} file accompanying this run takes the form of the default \texttt{standard\_run\_star\_extras.inc} file, slightly modified to handle the merger and profile write-out.
In particular, in the \texttt{extras\_check\_model} function, we add the following lines to initiate accretion at the proper time specified in the \texttt{inlist\_project} file:

\begin{lstlisting}[language=Fortran]
if (s% star_age >= s% x_ctrl(2) .and. s% x_ctrl(2) /= 0) then
    s% mass_change = s% x_ctrl(1)
end if
\end{lstlisting}

Additionally, in the function \texttt{extras\_finish\_step}, we add the following lines to force a write-out of the stellar profile at values of $\log L$ close to multiples of $0.1$.

\begin{lstlisting}[language=Fortran]
f = s% x_integer_ctrl(1)
s% xtra(1) = s% log_surface_luminosity

if ((floor(f * s% xtra_old(1)) - floor(f * s% xtra(1)) .ne. 0)) then
    s% need_to_update_history_now = .true.
    s% need_to_save_profiles_now = .true.
endif
\end{lstlisting}

In Section \ref{time}, we briefly discuss the usage of the \texttt{create\_merger\_model} feature to confirm the validity of modeling a MS+MS merger as a surface accretion event onto the primary.
Specifically, we have included in \texttt{inlist\_project} the option \texttt{write\_model\_with\_profile = .true.} for the $M=1.50$ $M_\odot$ and $M=1.00$ $M_\odot$ models, and have passed the saved model files at the desired time of accretion to \texttt{saved\_model\_for\_merger\_1} and \texttt{saved\_model\_for\_merger\_2}.

In Section \ref{clump}, we examine the period spacing for stars undergoing helium core burning on the red clump.
As the helium flash is a very difficult stage of evolution to model numerically, we include the following two lines in the \texttt{\&controls} section of \texttt{inlist\_project} in order to prevent the timestep from becoming prohibitively small:

\begin{lstlisting}
    use_dedt_form_of_energy_eqn = .true.
    convergence_ignore_equL_residuals = .true.
\end{lstlisting}

\bsp	
\label{lastpage}
\end{document}